\documentstyle[a4,epsfig,rotate,rotating,12pt]{article}

\newcommand{\vus}{\mbox{$|{\rm V}_{us}|$}}

\newcommand{\epem}{\mbox{${\rm e}^+{\rm e}^-$}}

\newcommand{\tkpppz}{\mbox{$\tau^- \!\!\rightarrow\!\! \,\nu_\tau\, {\rm K}^- \pi^0 \pi^0 \pi^0 $}}
\newcommand{\tkkppz}{\mbox{$\tau^- \!\!\rightarrow\!\! \,\nu_\tau\, {\rm K}^- {\rm K}^0 \pi^0 \pi^0$}}

\newcommand{\ekp}{\mbox{$\epsilon_{{\rm K}^-{\pi}^0}$}}
\newcommand{\ekk}{\mbox{$\epsilon_{{\rm K}^-{\rm K}^0}$}}
\newcommand{\ek}{\mbox{$\epsilon_{{\rm K}^-}$}}
\newcommand{\ekz}{\mbox{$\epsilon_{{\rm K}^-{\rm h}^0}$}}
\newcommand{\ekzz}{\mbox{$\epsilon_{{\rm K}^-\ge 2 {\rm h}^0}$}}
\newcommand{\eknz}{\mbox{$\epsilon_{{\rm K}^-\ge 0 {\rm h}^0}$}}
\newcommand{\epnz}{\mbox{$\epsilon_{{\pi}^-\ge 0 {\rm h}^0}$}}
\newcommand{\ekoz}{\mbox{${\epsilon}_{{\rm K}^-\ge 1 {\rm h}^0}$}}
\newcommand{\bk}{\mbox{${\rm B}_{{\rm K}^-}$}}
\newcommand{\bkz}{\mbox{${\rm B}_{{\rm K}^-{\rm h}^0}$}}

\newcommand{\bkzz}{\mbox{${\rm B}_{{\rm K}^-\ge 2 {\rm h}^0}$}}
\newcommand{\bknz}{\mbox{${\rm B}_{{\rm K}^-\ge 0 {\rm h}^0}$}}
\newcommand{\bpnz}{\mbox{${\rm B}_{{\pi}^-\ge 0 {\rm h}^0}$}}
\newcommand{\bkp}{\mbox{${\rm B}_{{\rm K}^-{\pi}^0}$}}
\newcommand{\bkk}{\mbox{${\rm B}_{{\rm K}^-{\rm K}^0}$}}

\newcommand{\np}{\mbox{$\tau^- \!\!\rightarrow\!\! \,\nu_\tau\, \pi^-$}}

\newcommand{\nptz}{\mbox{$\tau^- \!\!\rightarrow\!\! \,\nu_\tau\, \pi^- \le 2\pi^0$}}

\newcommand{\nppz}{\mbox{$\tau^- \!\!\rightarrow\!\! \,\nu_\tau\, \pi^- \pi^0 $}}
\newcommand{\npkz}{\mbox{$\tau^- \!\!\rightarrow\!\! \,\nu_\tau\, \pi^- {\rm K}^0$}}
\newcommand{\npppz}{\mbox{$\tau^- \!\!\rightarrow\!\! \,\nu_\tau\, \pi^- \pi^0 \pi^0 $}}
\newcommand{\npeta}{\mbox{$\tau^- \!\!\rightarrow\!\! \,\nu_\tau\, \pi^- \eta +\ge 0 \pi^0$}}
\newcommand{\nppppz}{\mbox{$\tau^- \!\!\rightarrow\!\! \,\nu_\tau\, \pi^- \pi^0 \pi^0 \pi^0$}}
\newcommand{\tpkkz}{\mbox{$\tau^- \!\!\rightarrow\!\! \,\nu_\tau\, \pi^- {\rm K}^0 {\rm K}^0$}}
\newcommand{\tpkpz}{\mbox{$\tau^- \!\!\rightarrow\!\! \,\nu_\tau\, \pi^- {\rm K}^0 \pi^0$}}

\newcommand{\tp}{\mbox{$\tau^- \!\!\rightarrow\!\! \,\nu_\tau\, \pi^-$}}
\newcommand{\tk}{\mbox{$\tau^- \!\!\rightarrow\!\! \,\nu_\tau\, {\rm K}^-$}}
\newcommand{\kl}{\mbox{${\rm K}^- \!\!\rightarrow\!\! \,\bar{\nu}_l\, l^-(\gamma)$}}
\newcommand{\tknz}{\mbox{$\tau^- \!\!\rightarrow\!\! \,\nu_\tau\, {\rm K}^- \ge 1 {\rm h}^0$}}
\newcommand{\thnnz}{\mbox{$\tau^- \!\!\rightarrow\!\! \,\nu_\tau\, {\rm h}^- \ge 0 {\rm h}^0$}}
\newcommand{\tknnz}{\mbox{$\tau^- \!\!\rightarrow\!\! \,\nu_\tau\, {\rm K}^- \ge 0 {\rm h}^0$}}
\newcommand{\tpnnz}{\mbox{$\tau^- \!\!\rightarrow\!\! \,\nu_\tau\, \pi^- \ge 0 {\rm h}^0$}}
\newcommand{\tkpz}{\mbox{$\tau^- \!\!\rightarrow\!\! \,\nu_\tau\, {\rm K}^- \pi^0 $}}
\newcommand{\tkkz}{\mbox{$\tau^- \!\!\rightarrow\!\! \,\nu_\tau\, {\rm K}^- {\rm K}^0$}}

\newcommand{\tpzzz}{\mbox{$\tau^- \!\!\rightarrow\!\! \,\nu_\tau\, \pi^- \ge 0 {\rm h}^0$}}
\newcommand{\tkzzz}{\mbox{$\tau^- \!\!\rightarrow\!\! \,\nu_\tau\, {\rm K}^- \ge 0 {\rm h}^0$}}
\newcommand{\tkzzzbf}{\mbox{${\bf \tau^- \!\!\rightarrow\!\! \,\nu_\tau\, {\rm K}^- \ge 0 {\rm h}^0}$}}

\newcommand{\tkz}{\mbox{$\tau^- \!\!\rightarrow\!\! \,\nu_\tau\, {\rm K}^- {\rm h}^0$}}
\newcommand{\tkzz}{\mbox{$\tau^- \!\!\rightarrow\!\! \,\nu_\tau\, {\rm K}^- \ge 2 {\rm h}^0$}}

\newcommand{\tketa}{\mbox{$\tau^- \!\!\rightarrow\!\! \,\nu_\tau\, {\rm K}^- \eta$}}
\newcommand{\nkppz}{\mbox{$\tau^- \!\!\rightarrow\!\! \,\nu_\tau\, {\rm K}^- \pi^0 \pi^0 $}}

\newcommand{\nkkpz}{\mbox{$\tau^- \!\!\rightarrow\!\! \,\nu_\tau\, {\rm K}^- {\rm K}^0 \pi^0$}}

\newcommand{\tkpp}{\mbox{$\tau^- \!\!\rightarrow\!\! \,\nu_\tau\, {\rm K}^-\pi^-\pi^+$}}

\newcommand{\tkppzz}{\mbox{$\tau^- \!\!\rightarrow\!\! \,\nu_\tau\, {\rm K}^-\pi^-\pi^+ \ge 0 \pi^0$}}

\newcommand{\tmu}{\mbox{$\tau^- \!\!\rightarrow\! \mu^- {\overline{\nu}}_\mu \,\nu_\tau$}} 
 
\newcommand{\dimuon}{\mbox{${\rm e}^+{\rm e}^- \!\!\rightarrow\! \mu^+\mu^-$}} 
 
\newcommand{\eeeemm}{\mbox{${\rm e}^+{\rm e}^- \!\!\rightarrow\! {\rm e}^+ {\rm e}^- \mu^+\mu^-$}} 
 
\newcommand{\eett}{\mbox{${\rm e}^+{\rm e}^- \!\!\rightarrow\! \tau^+\tau^-$}}

\newcommand{\de}{\mbox{${\rm d}E/{\rm d}x$}}
\newcommand{\dmeas}{\mbox{$D_{\rm meas}$}}
\newcommand{\dpred}{\mbox{$D_{\rm pred}$}}

\newcommand{\sbeta}{\mbox{$s(\beta)$}}   
   
\newcommand{\sres}{\mbox{$s_{res}$}}
\newcommand{\fphi}{\mbox{$f(\phi)$}}

\newcommand{\etal}{{\it et al.}}

\newcommand{\mh}{\mbox{hadronic ${\rm Z}^0\rightarrow q\bar{q}$}}
\begin{document}
\sloppy
\begin{titlepage}
\begin{center}{\large   EUROPEAN ORGANIZATION FOR NUCLEAR RESEARCH
}\end{center}\bigskip
\begin{flushright}
       CERN-EP-2000-091   
\\
       OPAL PR315
\\ 
       6th July, 2000
\end{flushright}
\bigskip\bigskip\bigskip\bigskip\bigskip
\begin{center}{\huge\bf   
   A Study of One-Prong \\
   Tau Decays with a Charged Kaon \\
       \  \\
}\end{center}\bigskip\bigskip
\begin{center}{\LARGE The OPAL Collaboration
}\end{center}\bigskip\bigskip
\bigskip\begin{center}{\large  Abstract}\end{center}
 
 From an analysis of the ionisation energy loss
 of charged particles selected from
 110326 \eett\ candidates recorded by the OPAL
 detector at \epem\ centre-of-mass energies near the ${\rm Z}^0$
 resonance,
 we determine the one-prong tau decay branching ratios:
 \begin{eqnarray*}
 {\rm Br}(\tkzzz) & = & 1.528 \pm 0.039 \pm 0.040 \%
 \nonumber \\[2mm]
 {\rm Br}(\tk)    & = & 0.658 \pm 0.024 \pm 0.029 \%
 \nonumber
 \end{eqnarray*}
 where the ${\rm h}^0$ notation refers to
 a $\pi^0$, an $\eta$, a ${\rm K}^0_{\rm S}$, or a ${\rm K}^0_{\rm L}$,
 and where the first
 uncertainty is statistical and the second is systematic.
 
\bigskip\bigskip\bigskip\bigskip
\bigskip\bigskip
\begin{center}{\large
(Submitted to European Physics Journal {\bf C})\\
}\end{center}
\end{titlepage}
\begin{center}{\Large        The OPAL Collaboration
}\end{center}\bigskip
\begin{center}{
G.\thinspace Abbiendi$^{  2}$,
K.\thinspace Ackerstaff$^{  8}$,
C.\thinspace Ainsley$^{  5}$,
P.F.\thinspace Akesson$^{  3}$,
G.\thinspace Alexander$^{ 22}$,
J.\thinspace Allison$^{ 16}$,
K.J.\thinspace Anderson$^{  9}$,
S.\thinspace Arcelli$^{ 17}$,
S.\thinspace Asai$^{ 23}$,
S.F.\thinspace Ashby$^{  1}$
% ...
}\end{center}\bigskip
\bigskip
$^{  1}$School of Physics and Astronomy, University of Birmingham,
Birmingham B15 2TT, UK
\newline
$^{  2}$Dipartimento di Fisica dell' Universit\`a di Bologna and INFN,
I-40126 Bologna, Italy
\newline
$^{  3}$Physikalisches Institut, Universit\"at Bonn,
D-53115 Bonn, Germany
\newline
% ...
$^{  5}$Cavendish Laboratory, Cambridge CB3 0HE, UK
\newline
% ...
$^{  8}$CERN, European Organisation for Nuclear Research,
CH-1211 Geneva 23, Switzerland
\newline
$^{  9}$Enrico Fermi Institute and Department of Physics,
University of Chicago, Chicago IL 60637, USA
\newline
% ...
$^{ 16}$Department of Physics, Schuster Laboratory, The University,
Manchester M13 9PL, UK
\newline
$^{ 17}$Department of Physics, University of Maryland,
College Park, MD 20742, USA
\newline
% ...
$^{ 22}$Department of Physics and Astronomy, Tel Aviv University,
Tel Aviv 69978, Israel
\newline
$^{ 23}$International Centre for Elementary Particle Physics and
Department of Physics, University of Tokyo, Tokyo 113-0033, and
Kobe University, Kobe 657-8501, Japan
\newline
% ...
Budapest, Hungary.

\section{Introduction}
\leavevmode\indent
Precise studies of the decays of the tau lepton have been
made possible by the availability of large, low background
tau-pair samples, such as those produced in \epem\
collisions at CESR and 
at LEP I.  This paper reports on an analysis of one-prong
tau decay modes containing a charged kaon, 
using the complete set of data
collected by the OPAL experiment between 1990 and 1995
at \epem\ centre-of-mass energies near the ${\rm Z}^0$ 
resonance.
The excellent charged particle identification capability of the OPAL
detector is exploited to 
obtain precise
measurements of the \tk\ and \tkzzz\ branching ratios, where
the charge conjugate decays are implied in these interactions and
throughout this paper.
The ${\rm h}^0$ notation refers to
a $\pi^0$, an $\eta$, or a ${\rm K}^0$, where
the ${\rm K}^0$ can be either 
a ${\rm K}^0_{\rm S}$ or a ${\rm K}^0_{\rm L}$. 
 
Measurements of the \tk\ branching fraction may be used
to determine the charged kaon decay constant, $f_{\rm K}$,  and the
CKM matrix element, \vus.  In addition, precise measurements
of this branching ratio may also be used to set limits on physics 
beyond the Standard Model, such as charged Higgs effects
in tau decays~\cite{bib:meb}.
 
\section{The OPAL Detector}
\leavevmode\indent
The OPAL detector is described in detail in reference
\cite{bib:opaldet}.  The component of the detector most
important to this analysis is a large volume
jet chamber,
which measures the momentum and
specific ionisation energy loss, \de, 
of charged particles. The jet chamber provides the only
means of distinguishing between charged pions and kaons in the
momentum range of interest to this study.
 
The jet chamber is a cylinder
4 m long and 3.7 m in diameter, and is divided 
by cathode wire planes        
into 24 azimuthal sectors.  
Each sector contains one radial plane of anode wires, which are
staggered to resolve left-right ambiguities.
The chamber is
contained in a solenoidal 
magnetic field of $0.435$ T,
and is filled with an argon-methane-isobutane gas mixture at a pressure
of 4 atmospheres.  This arrangement provides an intrinsic 
transverse spatial hit
resolution of $\sigma_{xy} = 130$ $\mu$m, and a two-hit resolution
of $2.5$ mm.\footnote{A spherical coordinate
system is used, with the $+z$-axis in the direction of the circulating
electron beam.  The angle $\theta$ is defined as the
polar angle with respect to the $+z$-axis, and $\phi$ is defined
as the azimuthal angle measured from the $+x$-axis, which points
towards the centre of the LEP ring.}
In the barrel region of the
jet chamber ($|\cos{\theta}| < 0.68$),
the ionisation energy
loss of a charged
particle is sampled up to 159 times.
A truncated mean is formed by discarding
the highest $30\%$ of the measurements, 
resulting in a \de\
resolution of $3.1\%$ for isolated tracks in the 
chamber~\cite{bib:dedxref}.
 
Other tracking detectors include a high precision silicon
microvertex detector surrounding the beam pipe, a precision gas vertex
detector, and 
a layer of wire chambers with drift direction parallel to the
$z$-axis lying immediately outside the jet chamber. 
These $z$-chambers extend
to $|\cos{\theta}|=0.80$, which encompasses the acceptance
of this analysis,  and
accurately determine the polar angle of charged particles
traversing the central detector.
A lead-glass electromagnetic calorimeter and presampler
chambers are located outside the magnetic coil and jet chamber
pressure vessel.  
The electromagnetic calorimeter is primarily used in
this analysis to identify tau decays that include a
$\pi^0$ or a ${\rm K}^0$ in the final state.
The return yoke of the OPAL magnet is instrumented for
hadron calorimetery, and is surrounded by external muon chambers.
The hadron calorimeter and muon chambers are mainly
used in this analysis to select control samples containing
muons and to veto non-tau backgrounds in the
tau-pair candidate sample.
 
\subsection{Parameterisation of Ionisation Energy Loss}
\leavevmode\indent
 The mean ionisation
 energy lost per unit distance by a particle of charge
 $q$ and speed $\beta$
 traversing a gas
 is approximately
 described by the Bethe-Bloch
 equation \cite{bib:bethe,bib:fermi}:
\begin{eqnarray}
 D_{\rm pred} = -{{A_1 q^2}\over{\beta^2}} \left[A_2
              + \ln{\left(\gamma^2 \beta^2\right)}
               -\beta^2 -\delta(\beta)/2 \right] ,
\label{eqn:bethe}
\end{eqnarray}
 where
 $A_1$ and $A_2$ are constants which depend on the gas
 composition,
 and $\delta(\beta)$ is a function which describes polarisation
 effects in the gas.
 
\begin{figure}[p]
   \begin{center}
     \mbox{ \epsfxsize=15cm
            \epsffile[23 165 524 652]{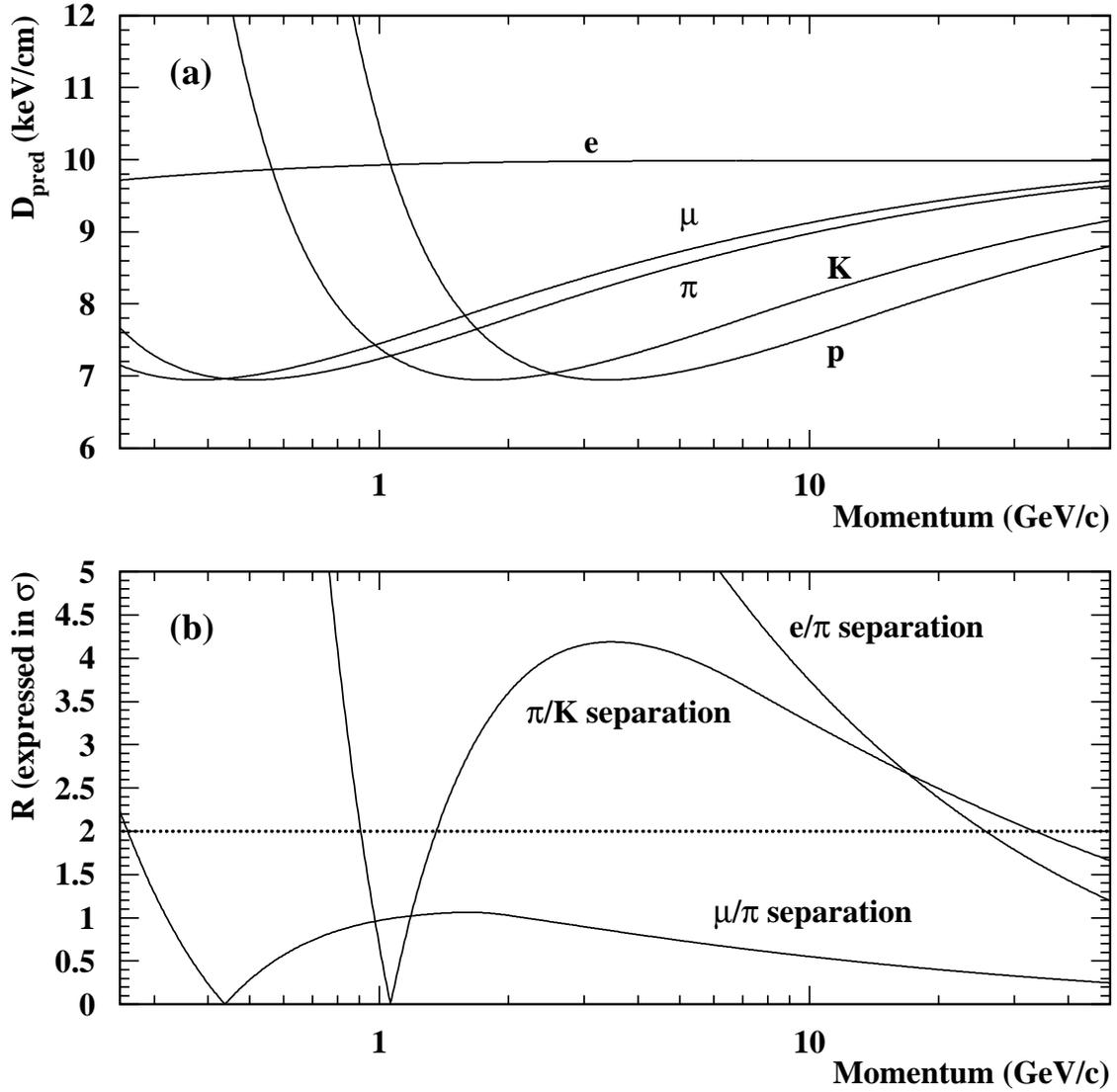} }
   \end{center}
 \vspace*{0.5cm}
 \caption[foo]{
 \label{fig:predict}
 (a) shows the ionisation energy loss \dpred\ as a function
 of the momentum for various particles in the OPAL jet chamber.
 (b) shows the resolution power ${\cal R}_{ij}$ 
 expressed in
 terms of the \de\ resolution $\sigma$,
 for various pairs of particle species $i$ and $j$.
 }
 \end{figure}

 The ionisation energy loss parameterisation used by OPAL, \dpred,
 is 
 a variation of 
 the Sternheimer-Peierls modification to the Bethe-Bloch
 formula \cite{bib:dedxref,bib:stern},
 and
 uses six input parameters.
 Figure \ref{fig:predict}(a) shows the dependence of \de\
 on the momentum of various particles in the OPAL jet chamber,
 as predicted by
 \dpred.
 Figure \ref{fig:predict}(b) shows the
 particle resolving power, ${\cal{R}}_{ij}$,
 versus momentum for various pairs of particle species $i$ and $j$, 
 where:
\begin{eqnarray}
{\cal{R}}_{ij} \equiv |D^i_{\rm pred} - D^j_{\rm pred}| \, / \, \sigma  ,
\nonumber
\end{eqnarray}
 and where $\sigma$ is the 
 resolution of the measured \de.
 Note that the OPAL jet chamber
 yields a pion/kaon separation of at least
 $2\sigma$ for particles between about 2 to 30 GeV/c.

The parameters used in the calculation of \dpred\ and
$\sigma$ 
are tuned to the measured energy loss of charged particles in \mh\
decays, yielding a \de\ parameterisation that is accurate enough
for nearly all analyses of OPAL data.
 
However, \mh\ decays tend to have a much higher track multiplicity than 
one-prong tau decays, leading to small
systematic differences between the measured \de\ in the two environments.
Although these differences
are less than one $\sigma$ in magnitude, they must be corrected
for accurate particle identification in one-prong tau decays.  
Based on a study of the measured energy loss of 
muons in \dimuon, \tmu, and \eeeemm\ control samples,
the following corrections are applied to the
the OPAL 
\de\ parameterisation, \dpred, and to the 
measured energy loss 
and its uncertainty, \dmeas\ and $\sigma$:
\begin{description}
\item[{\rm Multiplicative correction to \dpred\ :}]
      To obtain a parameterisation which 
      correctly predicts the measured 
      energy loss of particles from tau decays, \dpred\
      must be corrected with a multiplicative factor, $s(\beta)$.
      Parameterising $s(\beta)$ as a second
      order polynomial in $-\log(1-\beta^2)$ yields 
      satisfactory results.  
      The magnitude of the
      correction is of order $1\%$ for tracks in one-prong tau decays.
\item[{\rm Additive correction to \dmeas\ :}]
      The measured energy loss of charged
      particles in low-multiplicity events is found
      to depend strongly on 
      the azimuthal separation, $\phi$, between the track and
      anode plane in the jet chamber cell.
      To improve the \de\ resolution, this behaviour is
      corrected with a function, $f(\phi)$, with
      seven parameters tuned to the
      measured energy loss of \dimuon\ candidates.
      The magnitude of this correction is also of order $1\%$.
\item[{\rm Multiplicative correction to $\sigma$:}]
      To correctly
      predict the spread in the \de\ of charged particles
      from one-prong tau decays,
      the OPAL parameterisation of $\sigma$ must be
      corrected
      with a multiplicative factor, $s_{\rm res}\approx 0.9$.  
\end{description}
 
\section{Monte Carlo Generated Event Samples} 
\leavevmode\indent
For this analysis, tau lepton decays are simulated with
the 
KORALZ $4.0$ Monte Carlo generator and the 
TAUOLA $2.4$ decay package \cite{bib:koralz,bib:tauola}.
A Monte Carlo sample of $600000$ tau-pair
events is generated with
input branching ratios determined from world averages or theoretical
expectations~\cite{bib:pdg94}.
Background contributions from non-$\tau$ sources 
were evaluated using Monte Carlo samples obtained from
the following generators:
\mh\ events were simulated using 
JETSET 7.4~\cite{jetset}, \dimuon\
events using KORALZ,  
Bhabha events using BABAMC~\cite{babamc} and BHWIDE~\cite{bhwide},
and four-fermion events using 
VERMASEREN 1.01~\cite{verm} and FERMISV~\cite{fermisv}.

All Monte Carlo samples are passed through a detailed
simulation of the OPAL detector~\cite{bib:geant} and are
subjected to the same analysis chain as the data.
 
\section{Data Selection}
This analysis uses the full data set collected
by the OPAL detector
in the years 1990 through 1995 at \epem\ centre-of-mass energies
close to the ${\rm Z}^0$ resonance.  Only data for which
the tracking chambers
and the electromagnetic calorimeter were fully operational are retained.
The topology of \eett\ events is characterised by a pair of back-to-back,
narrow jets with low particle multiplicity.  Jets are defined in this
analysis by grouping tracks and electromagnetic clusters into
cones with a half-angle of $35^\circ$, where each cone is assumed
to contain the decay products of one of the tau leptons.
Tau-pair candidates are selected by requiring two low-multiplicity
jets
($N_{\rm track} \le 6$)
with an average polar angle of $|\cos{\theta_{\rm ave}}| < 0.68$.
Background events from other lepton pairs, \mh\ decays,
and four-fermion events
are reduced with cuts on the event topology and total visible
energy.  
These selections produce a sample of 110326 tau-pair candidates
with background
$f^{{\rm non-}\tau} = 1.56 \pm 0.10 \%$,
estimated by data control samples and Monte
Carlo background samples.  
The tau-pair selection procedure 
is described in detail in reference \cite{bib:john}.

One-prong tau decays are selected from this sample
by choosing tau decay cones containing only one well-reconstructed
charged track, consistent with coming from the origin.
Tracks in this sample are required to
have at least 40 jet chamber hits used in the
calculation
of the measured \de,
to have at least 3 hits in the $z-$chambers,
and to have reconstructed momentum between 2 and 50~GeV/c.
A sample of 143528 candidates is selected by this procedure.

The number of charged kaons in the sample is determined by maximising
a likelihood function based on \de. 
The \tkzzz\ branching ratio is then obtained by correcting the number
of kaons
for backgrounds and selection inefficiencies.  
As a cross-check of the general analysis procedure,
the \tpzzz\ branching
ratio is also determined in an analogous fashion, and compared
to the world average.
 
\section{Inclusive \tkzzzbf\ Branching Ratio}
\leavevmode\indent
The tracks in the one-prong tau decay candidate
sample consist of electrons, muons,
pions, and kaons from both
tau decays and non-tau sources.
To determine the number of charged kaons in the sample, a
maximum likelihood fit to the ionisation energy loss distribution
of the tracks in the sample is performed.
The likelihood function used in the fit is:
\begin{eqnarray}
 {\cal{L}} =
 \exp{\left[-{\textstyle {1\over2}}\,
 \left({{f^\prime_\mu-f_\mu}\over{\Delta f^\prime_\mu}}\right)^2
 \right]
 }
            \prod_{j = 1,N} \;
            \sum_{k = {\rm e},\mu,\pi,{\rm K}}
            f_k \; W^{kj} ,
\label{eqn:like1}
\end{eqnarray}
where
$W^{kj}$ is the \de\ weight of charged particle $j$
under particle hypothesis $k$,
\begin{eqnarray}
W^{kj} =    {1\over{\sqrt{2\pi} s(\beta_j) s_{\rm res} \sigma_{kj}}}
            \;
            \exp{
            \left[ - {{1}\over{2}}
            \left(
                { {D_{\rm meas}^j - f(\phi_j) - s(\beta_j) D_{\rm pred}^{kj}}
                \over
                {s(\beta_j) s_{\rm res} \sigma_{kj}} }
            \right)^2
            \right]
            } ,
\label{eqn:like2b}
\end{eqnarray}
and where
\begin{description}
\item[{\rm $f_k$ }] is the fraction of particle type $k$ in the sample,
where $k$ is either pion, kaon, muon,
or electron.  The values of $f_k$ are
constrained to be non-negative, and their sum is constrained to be 1.
 \item[{\rm $f^\prime_\mu$ and $\Delta f^\prime_\mu$}] are the
 fraction, and uncertainty on the fraction, of muons in the sample, as
 estimated by Monte Carlo generated events. The estimated muon fraction
 is corrected using information
 from data and Monte Carlo muon control samples that are selected using
 information from the OPAL muon chambers.
 This constraint is added to the fit to help distinguish between pions and
 muons in the sample, which have quite similar \de.
\item[{\rm $N$ }] is the total number of particles
in the sample.
\item[{\rm $D_{\rm meas}^j$ }] is the measured \de\ of the $j^{\rm th}$
charged particle.
\item[{\rm $D_{\rm pred}^{kj}$ }] is the predicted \de\
for charged particle $j$, calculated with the
OPAL parameterisation under particle hypothesis $k$,
as obtained from the measured \de\ of charged particles in \mh\ events.
\item[{\rm $\sigma_{kj}$ }] is the \de\ uncertainty,
calculated
using the OPAL parameterisation,
as obtained from the measured \de\ of \mh\ events.
\item[{\rm \sres }] is the multiplicative correction
to $\sigma_{kj}$,
as determined from the  
one-prong control samples as described in Section~$2.1$.
\item[{\rm \sbeta }] is the $\beta$ dependent multiplicative correction
to \dpred,
as determined from the one-prong control samples.
\item[{\rm \fphi }] is the $\phi$ dependent correction to the
measured \de, as obtained from the one-prong control samples.
\end{description}

Efforts are made to obtain a \de\
parameterisation for the tau decay environment
that is optimal for many particle species
over a wide range of momenta.  However, it is possible that
the \de\ corrections described
in Section~$2.1$ may be somewhat
more (or less) optimal for pions than they are for
kaons in the momentum range of interest.
Thus, to correct for any possible
species-dependent quality differences
in the parameterisation of \de,
an
extra factor, $C_{\rm K}$, is allowed to multiply
the kaon predicted energy loss, $D_{\rm pred}^{\rm K}$, and is
allowed to vary freely in the fit using the likelihood
function found in Equation~\ref{eqn:like1}.
The value of $C_{\rm K}$ returned by the fit is
$C_{\rm K} = 0.9943 \pm 0.0009$.

Independent likelihood fits are performed in 13 momentum bins of variable
size in the range of 2 to
50 GeV/c.
A test of the fit with Monte Carlo generated
samples verifies that the resulting estimates for the
kaon fraction are efficient 
and unbiased.

\begin{figure}[ht]
   \begin{center}
     \mbox{ \epsfxsize=15cm
            \epsffile[23 165 524 652]{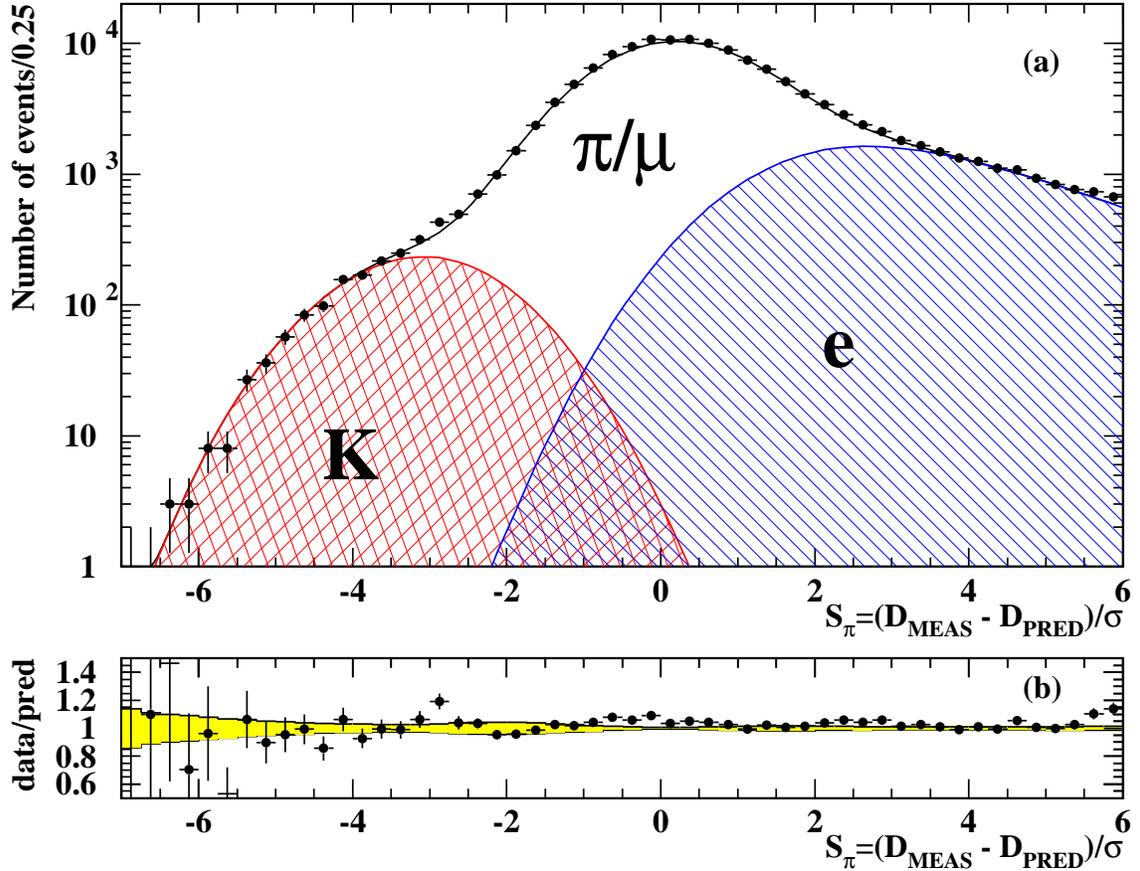} }
   \end{center}
 \vspace{-0.5cm}
 \caption[Stretch \de\ distribution of tracks in the \thnnz\ candidate sample
]{
 \label{fig:fit1b}
  (a) is the stretch \de\ distribution under a pion hypothesis for tracks
  in the data \thnnz\ candidate sample (points).
  The overlaid curves are the
  predicted distributions for kaons, pions, muons, and electrons in the
  sample.  The normalisation of the curves is obtained from the results
  of the likelihood fit described in Section~$5$.
   (b) is the distribution of the
   data points in plot (a) divided by the predicted
   distribution.  The shaded area represents the approximate one
   sigma \de\ systematic uncertainty envelope on the predicted distribution.
   The $\chi^2$ per degree of freedom between the data and a line
   centred at $y=1$, taking into account both statistical and
   \de\ systematic uncertainties, is $36.5/47$.
 }
 \end{figure}

\begin{figure}[ht]
   \begin{center}
     \mbox{ \epsfxsize=15cm
            \epsffile[23 165 524 652]{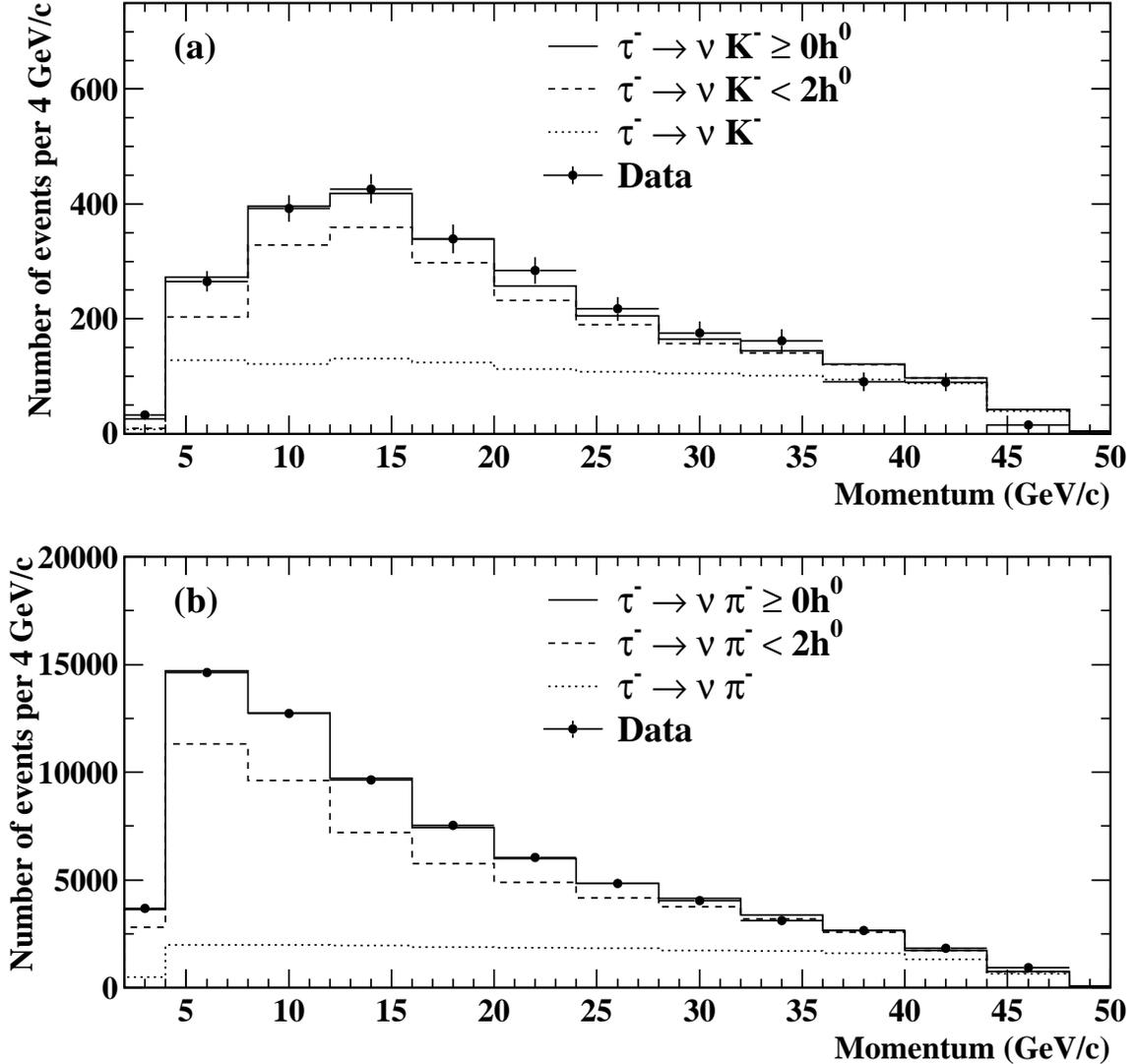} }
   \end{center}
 \vspace{-0.5cm}
 \caption[Momentum distribution of charged hadrons in the one-prong tau decay
 sample]{
 \label{fig:mom}
  (a) and (b) are the
  momentum distributions of charged kaons and pions, respectively,
  in the data one-prong tau decay
  sample (points), as obtained from the results
  of the \de\ likelihood fit described
  in Section~$5$.  The histograms are the momentum distributions
  predicted by Monte Carlo generated \tknnz\ and \tpnnz\ events.
  The overall normalisation of the predicted distributions
  comes from the results of the \de\ likelihood fit, while
  the relative normalisation of the exclusive decay modes
  contributing to the sample is taken from the relevant world
  average branching ratios appearing in reference \cite{bib:pdg}.
 }
 \end{figure}

The number of charged kaons and pions
found in the sample, summed over all momentum bins,
is given in
Table~\ref{tab:sum1b}, and
the stretch \de\ distribution of tracks in all momentum
bins of the sample is shown
in Figure~\ref{fig:fit1b}.  
Stretch energy loss,
$S_i$, is defined as:
\begin{eqnarray}
S_i = (D_{\rm meas} - D^i_{\rm pred})/\sigma ,
\nonumber
\end{eqnarray}
where $i$ is the particle hypothesis used to calculate the predicted
\de.
The
normalisation of the predicted distributions of the kaons, pions,
muons, and
electrons in Figure~\ref{fig:fit1b} 
is obtained from the results of the likelihood
fit.
The momentum distributions of the charged kaons and pions in the sample, as
estimated by the likelihood fit, are shown in Figure \ref{fig:mom}.
 
\subsection{The Branching Ratio Calculation}
\leavevmode\indent
To determine the \tkzzz\ branching ratio, the number of
charged kaons found by the \de\ likelihood fit
in the one-prong candidate sample
is corrected for backgrounds and selection inefficiencies:
\begin{eqnarray}
\bknz \equiv {\rm Br}(\tkzzz) = {
{N^{\rm K}_{\rm TOTAL} - N_{\rm bkgnd}^{\rm K} }\over
{\eknz N_\tau (1-f^{{\rm non}-\tau})}},
\label{eqn:brcalc}
\end{eqnarray}
where
\begin{description}
\item[$N_\tau$] is the number of tau decay
candidates.  There were 220652 tau decay candidates recorded
in the barrel region of the OPAL detector between the years
1990 and 1995.
\item[$f^{{\rm non-}\tau}$] is the estimated background from non-$\tau$
sources in the tau decay candidates
($f^{{\rm non-}\tau} = 1.56\pm0.10\%$).
\item[$N^{\rm K}_{\rm TOTAL}$] is the total number of kaons in the one-prong
tau decay sample, as estimated by the likelihood fit.  The number
of kaons summed over all momentum bins in the sample is shown in
Table~\ref{tab:sum1b}.
\item[$N_{\rm bkgnd}^{\rm K}$] is the number of background
kaons in the one-prong
tau decay sample, as estimated by Monte Carlo generated
tau decay and \mh\ samples.  The number
of background kaons summed over all momentum bins in the sample is shown in
Table~\ref{tab:sum1b}.
\item[\eknz] is the efficiency for \tknnz\ decays in the
tau-pair candidate sample to contribute kaons to the one-prong
tau decay sample, and is shown in Table~\ref{tab:sum1b}.
The efficiency has been corrected
for biases introduced by the tau-pair selection procedure, and
the uncertainty on the efficiency includes the systematic
uncertainty arising from this correction.
The bias corrections 
are all consistent with 1 to within one standard deviation
of their Monte Carlo
statistical uncertainties.
\end{description}

The result of the branching ratio calculation is:
\begin{eqnarray}
\bknz & = & 1.528 \pm 0.039 \pm 0.040 \%,
\nonumber
\end{eqnarray}
where the first uncertainty is statistical and the second is
systematic.
A summary of the systematic uncertainties, all of which
will be described in subsequent sections, is
shown in Table~\ref{tab:sum2b}.

As a cross-check of the general analysis procedure,
we
also determine the \tpzzz\ branching ratio.
Thus, in a similar manner to the
\tkzzz\ branching ratio calculation, the total number
of charged pions found in the one-prong tau decay
sample by the likelihood
fit is corrected for backgrounds and the \tpzzz\ efficiency
to yield
$\bpnz = 48.26 \pm 0.25 \%$,
where the uncertainty is statistical only.
The linear correlation coefficient between the
OPAL
\tkzzz\ and \tpzzz\ branching ratios is approximately $-20\%$.
The OPAL \tpzzz\ branching ratio
is consistent with the world average, derived from
the branching ratios of the tau to the various exclusive final
states that include a charged pion listed in reference \cite{bib:pdg},
$\bpnz  =  48.36\pm0.23 \%$,
and thus we conclude that biases introduced by the analysis
procedure are negligible.

\subsection{\de\ Systematic Studies}
\leavevmode\indent
A significant source of systematic uncertainty in the measurement
of the inclusive branching ratio
is the uncertainty in the parameterisation
of the predicted energy loss.
To assess this systematic,
we determine the sensitivity of the likelihood estimates of the
number of kaons within
the data sample
to the uncertainties in the
\de\ correction factors obtained from the one-prong control samples.
The likelihood function shown in Equation~\ref{eqn:like1}
is therefore modified
such that:
\begin{eqnarray}
 {\cal{L}}^\prime = \exp{(-{\textstyle {1\over2}}\,
{\bf (n^\prime-n)^T V^{-1} (n^\prime-n)})} \; {\cal{L}} ,
\nonumber
\end{eqnarray}
where
\begin{description}
\item[${\bf n^\prime}$] is a vector containing the
central values of the
\de\ correction factors that describe the functions $f(\phi)$,
$s(\beta)$, and $s_{\rm res}$,  where the central values are
as obtained from the fits to the one-prong control samples.
\item[${\bf n}$] is a vector containing the
\de\ correction factors used in the likelihood fit.
\item[${\bf V}$] is the covariance matrix
for the \de\ correction factors,
as obtained from the fits to the one-prong control samples.
\end{description}

A `prior-belief' method is used to assess the systematic uncertainty;
in the
first iteration of the fit, the correction
factors are allowed to vary, and the returned values are
found to be consistent with the values obtained from the control samples.
In the second iteration, the likelihood fit is repeated, this time 
keeping the
\de\ correction
factors fixed to the values from the first iteration.  The
systematic uncertainty in $N^{\rm K}_{\rm TOTAL}$ (and $N^{\rm \pi}_{\rm TOTA
L}$) due to the
parameterisation of \de\ is then obtained from the
square root of the
quadrature difference of the fit uncertainties in $N^{\rm K}_{\rm TOTAL}$
($N^{\rm \pi}_{\rm TOTAL}$)
from the two iterations, and is shown in
Table~\ref{tab:sum1b}.


\subsection{Efficiency Estimation}
\leavevmode\indent
The efficiencies for kaons and pions from signal events
in the tau-pair candidate sample to enter
the data sample are estimated
using signal events generated with the KORALZ $4.0$
Monte Carlo generator and the
TAUOLA $2.4$ decay package, as described in Section~3.
The efficiency for \tkzzz\ decays in the
tau-pair candidate sample to contribute
to the one-prong sample, \eknz, is calculated via:
\begin{eqnarray}
 \eknz = f \ek + (1-f)\ekoz ,
\nonumber
\end{eqnarray}
where $f$ is the ratio of
the exclusive \tk\ branching ratio to the inclusive \tkzzz\
branching ratio, as obtained from OPAL data (see Section~$6.2$).
The \ekoz\ efficiency is calculated
from the average of the relevant efficiencies
appearing in Table~\ref{tab:br}, weighted by the
associated world average branching ratios.  
The inclusive and exclusive branching ratios determined
by this analysis are all less than $15\%$ correlated with the
world average branching ratios used in this efficiency calculation.
All efficiencies have been corrected
for biases introduced by the tau-pair selection procedure, and,
in addition,
the Monte Carlo efficiency estimates for the signal events
to pass the jet chamber and $z$-chamber hit requirements are
corrected using a control sample of data one-prong tau decays. 
The uncertainty on each efficiency includes the systematic
uncertainties arising from these corrections.

The efficiency for \tpzzz\ decays in the
tau-pair candidate sample to contribute
to the one-prong sample, \epnz, is calculated
using the same procedure, except that the ratio of the \tp\ branching
ratio relative to the \tpzzz\ branching ratio is taken from world averages.

The Monte Carlo modelling of the probability for photons to convert to
an \epem\ pair
in the OPAL detector material
has been found by previous studies of data control samples
to be correct to within
an absolute uncertainty of
about $\pm0.5\%$~\cite{bib:taumu}.  To assess the systematic 
uncertainty in the branching fraction associated with 
this modelling, 
we determine the change in the
branching fraction when 
the efficiency for each exclusive
final state is changed by $+0.005$ for each associated photon in the final state;
thus the efficiency for \tkpz\ decays is changed by $2\times0.005=0.01$,
and the efficiency for \nkppz\ decays is changed by $4\times0.005=0.02$.
We also determine the change in the
branching fraction when 
the efficiency for each exclusive
final state is changed by $-0.005$ for each associated photon in the final state.
The resulting systematic uncertainties in the branching fractions are shown
in Table~\ref{tab:sum2b}.
 
The variation of the branching ratio due
to alternative intermediate resonant structure scenarios for the \nkppz\
final state is assessed using efficiency estimates from
signal events generated by the modified version of
TAUOLA~$2.4$ through non-resonant production only.
The efficiency calculated from this
sample for \nkppz\ to contribute to the data candidates is
within $0.4\sigma$ of the central value, where $\sigma$
represents the combined Monte Carlo
statistical uncertainties.
Since the two different scenarios produce consistent
results for the efficiency, and since there
is no {\it a priori} reason to expect
a strong efficiency dependence on the intermediate
resonant structure, no systematic uncertainty for this effect is
assigned.

In a similar fashion, the variation of the branching ratio due
to alternative intermediate resonant structure scenarios for the \nkkpz\
final state is explored using events generated
through non-resonant production only.  The \nkkpz\ efficiency
calculated from these events is
within $0.6\sigma$ of the central value, thus no additional systematic
uncertainty is assigned.

Since the observed momentum distributions for both the pions
and the kaons in the sample are well described by the Monte
Carlo generated events, 
no systematic uncertainty is assigned for
possible error in either the estimate of the total efficiency, or
the momentum dependence of the efficiency estimates, since
it is apparent that such errors are small relative to the other
systematic uncertainties, as listed in Table~\ref{tab:sum2b}.

\subsection{Background Estimation}
\leavevmode\indent
The number of
background kaons in $N^{\rm K}_{\rm TOTAL}$ is estimated by
applying the same selection criteria to
Monte Carlo samples of \mh\ decays and three-prong tau decays.
From the number of selected events that contain kaons, the
estimated number of background kaons is derived, as summarised in
Table~\ref{tab:sum1b}.  There are very few background kaons in
$N^{\rm K}_{\rm TOTAL}$, most of which result from the small amount of
low-multiplicity \mh\ events contaminating the tau-pair sample.
Due to this small contamination, the systematic uncertainty associated
with the background correction is neglible.

The number of
background pions in $N^{\pi}_{\rm TOTAL}$ is estimated in a
comparable manner.

\section{Exclusive Branching Ratios}
To distinguish between one-prong tau final states with and without
accompanying neutral particles,
information from the electromagnetic
calorimeter is used.
Tau decays that include neutral particles tend to
have a
larger relative electromagnetic
energy deposition, $E/p$,
associated with the
tau decay cone
than tau decays
without accompanying neutrals.  Such decays also
tend to have a larger number of associated
clusters,
$N_{\rm clus}$, in the electromagnetic
calorimeter than tau decays without neutrals.
To ensure accurate modelling of the $E/p$ and $N_{\rm clus}$
distributions of
one-prong tau decays to charged kaons,
corrections to the Monte Carlo
simulation are obtained from a data \tpzzz\ control sample.
The following sections describe
these corrections, and how they are applied in the
determination of the \tk\ branching ratio.

\subsection{Corrections to the Modelling of $E/p$ and $N_{\rm clus}$}
The two-dimensional $(p,E/p)$ and $(p,N_{\rm clus})$ distributions of
data \tpzzz\ decays
are used to determine corrections
to the Monte Carlo modelling of the
$E/p$ and $N_{\rm clus}$ distributions of \tkzzz\ decays.
To this end, the data one-prong tau decay sample is
divided into 13 bins of
momentum between
2 to 50 GeV/c and
6 bins of $E/p$, and
the number of charged pions and kaons within each bin is determined
by maximising
the \de\ likelihood
function shown in Equation \ref{eqn:like1}.   After background correction, 
this procedure yields the
$(p,E/p)$ distributions of \tpzzz\ and \tkzzz\ decays.
The data one-prong sample is also
divided into 13 bins of
momentum
and
4 bins of $N_{\rm clus}$, and the $(p,N_{\rm clus})$ distributions
of \tpzzz\ and \tkzzz\ decays are determined in a
comparable manner.

\begin{figure}[ht]
    \begin{center}
      \mbox{ \epsfxsize=15cm
            \epsffile[23 165 524 652]{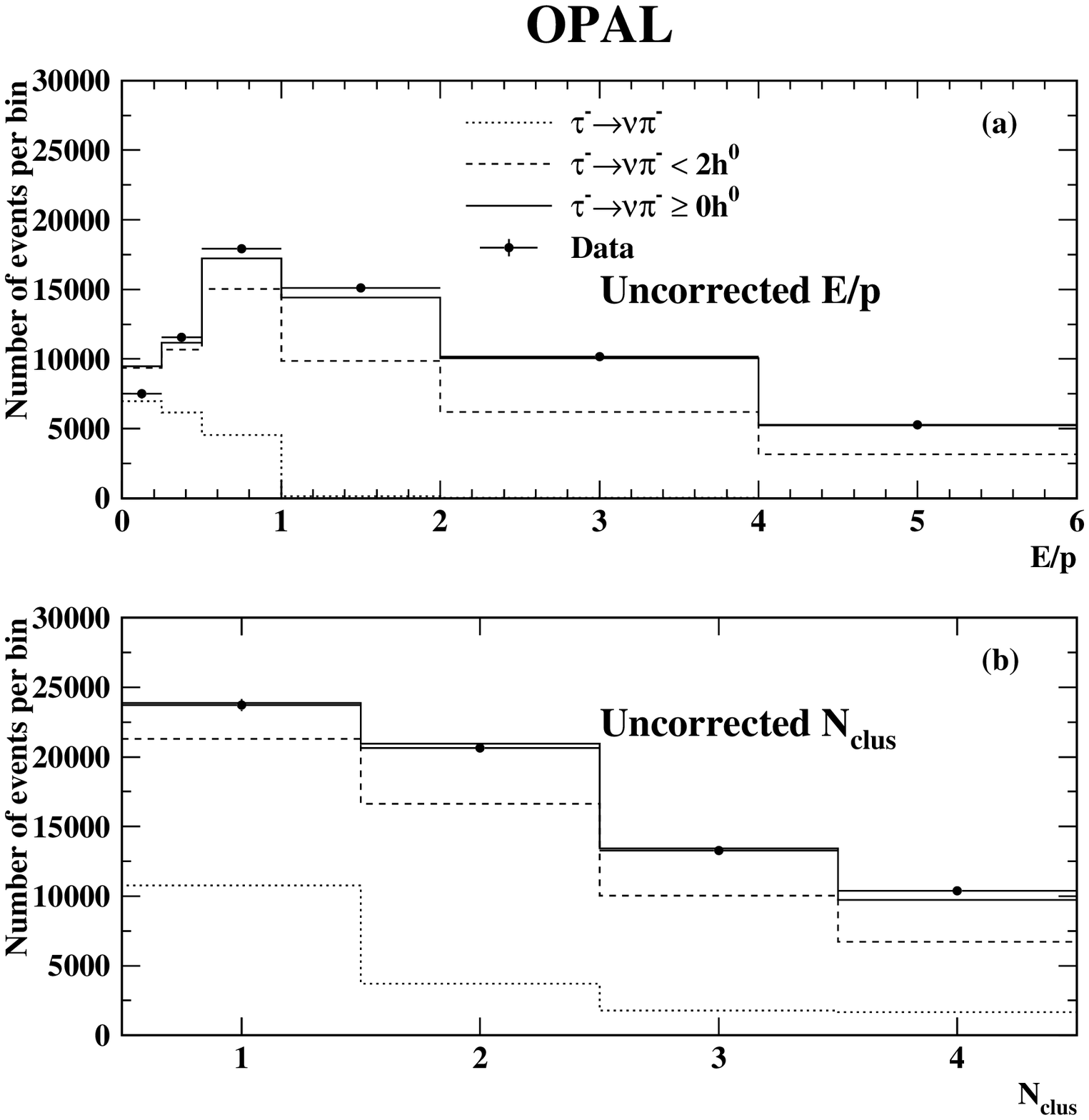} }
    \end{center}
  \caption[$E/p$ and $N_{\rm clus}$ distributions of \tpzzz\ decays]{
  \label{fig:pip3}
(a) and (b) are the uncorrected
$E/p$ and $N_{\rm clus}$ distributions, respectively, of
Monte Carlo generated \tpzzz\ decays (histograms), along with the
$E/p$ and $N_{\rm clus}$
distributions of data one-prong tau decays with a charged
pion in the final state (points).
The relative normalisations of the various exclusive tau decays that
contribute to the sample are taken from the relevant world average
branching ratios listed in reference \cite{bib:pdg}.
Backgrounds are on the order of $0.2\%$ or less and
are neglected in the plots.
The last bin in each of the distributions
contains overflow events.
  }
\end{figure}

The momentum
distribution of the charged pions in the one-prong tau decay sample,
as shown in Figure \ref{fig:mom},
appears to be adequately modelled by Monte Carlo generated \tpzzz\
events.
However, the $E/p$ and $N_{\rm clus}$ distributions, obtained
by summing the $(p,E/p)$  and $(p,N_{\rm clus})$ distributions
over momentum bins, are not as well modelled by the Monte Carlo, as
is shown in Figure \ref{fig:pip3}.

\subsubsection{Corrections to $E/p$}
Separate corrections are derived for the
Monte Carlo $E/p$ modelling of \tp, \nppz, and \npppz\
decays.  These three decay modes produce 95\% of the
charged pions found in the one-prong tau decay sample.
The $E/p$ corrections obtained for \tp\ are applied
to the Monte Carlo modelling of the $E/p$ of \tk, and,
in a
similar
fashion, the 
corrections to 
the $E/p$ distribution of 
\nppz\ (\npppz) decays 
are applied to
that of \tkz\ (\tkzz) decays.

The corrections to the \tp, \nppz, and \npppz\
Monte Carlo modelling of $E/p$ are obtained by first correcting
the number of charged pions within each $(p,E/p)$ bin
of the one-prong tau decay sample
for background pions from \mh\
events. The number of pions in each bin is also corrected
for the number of pions from tau sources other than
\tp, \nppz, and \npppz\ decays using world average
branching ratios and information from
Monte Carlo generated events.
The following $\chi^2$ function is then minimised:
\begin{eqnarray}
 \chi^2 = \sum_i \sum_j \left[
                       {{(N_{ij}^\pi-
                         \sum_l
                         N_\tau
                         (1-f^{{\rm non-}\tau}) \epsilon_l {\rm B}_l
                         \,\, s^{l}_{i} \,\, {S^l_{ij}}
                           \,\, {{\cal{S}}}_{j}^l
                           ) }
                          \over {\Delta N_{ij}^\pi}} \right]^2,
\label{eqn:chi1}
\end{eqnarray}
where the index $i$ runs over the momentum bins, the index $j$ runs
over the $E/p$ bins, and the index $l$ runs over \tp, \nppz, and
\npppz. Also:
\begin{description}
\item[$N_{ij}^\pi$] is the number of charged pions, after corrections
for non-\nptz\ background, found by the \de\
likelihood fit in the $(i,j)^{\rm th}$
bin of the $(p,E/p)$ distribution of one-prong tau decays.
\item[$\Delta N_{ij}^\pi$] is the statistical
uncertainty on $N_{ij}^\pi$, as obtained from the \de\
likelihood fit.
\item[{\rm $B_l$}] is the world average branching ratio of tau 
decay $l$, taken
from reference~\cite{bib:pdg}.
\item[{\rm $\epsilon_l$}] is the efficiency for events from tau
decay $l$ in the tau-pair candidate sample to contribute
to the one-prong tau decay sample.
\item[$S^l_{ij}$] is the probability, calculated
using Monte Carlo generated events, for tau decays of
type $l$ in the one-prong tau decay sample to contribute to the $(i,j)^{\rm t
h}$
bin of the $(p,E/p)$ distribution.
\item[${{\cal{S}}}_{j}^l$] is the correction to the
$j^{\rm th}$ bin of the Monte Carlo
$E/p$ distribution
of tau decays of type $l$.
The ${{\cal{S}}}_{j}^l$ are the only parameters allowed to float freely
in the fit.
\item[$s^l_{i}$] is a normalisation factor used to ensure that the
correction to $E/p$ does not affect the shape of the momentum distribution
of decays of type $l$, and is determined from the condition
$\sum_j s^l_i S^l_{ij} {{\cal{S}}}_{j}^l = P^l_i$, where $P^l_i$ is the
probability, as calculated from Monte Carlo generated
events, for tau decays of type $l$ to contribute to the
$i^{\rm th}$ bin of the momentum distribution.
\end{description}

\subsubsection{Corrections to $N_{\rm clus}$}
Corrections, ${{\cal{T}}}$, to the Monte Carlo modelling of $N_{\rm clus}$
of \tp, \nppz, and \npppz\
decays are obtained in a comparable manner to the method
used to obtain the
$E/p$ corrections.
Again,
the \tp, \nppz, and \npppz\ $N_{\rm clus}$
distributions are similar to those of \tk, \tkz, and \tkzz\ decays,
respectively.

The absolute linear correlation
coefficients between $N_{\rm clus}$ and $p$, and $N_{\rm clus}$ and
$E/p$ are less than $20\%$ in all classes of signal.
Thus, the $N_{\rm clus}$
and the $(p,E/p)$ distributions
are used as discriminators in this analysis, and
$N_{\rm clus}$ is treated as a variable uncorrelated to both
$p$ and $E/p$.

\subsection{The \tk\ Branching Ratio}
The \tk\ branching ratio, \bk,
is determined by a $\chi^2$ fit
to the $(p,E/P)$ and $N_{\rm clus}$ distributions
of one-prong tau decays with
a charged kaon, where the modelling of the distributions is corrected using
the results
of the previous section.  The following $\chi^2$ is minimised:
\begin{eqnarray}
 \chi^2 & = & \sum_j \sum_i \left[
                       {{N_{ij}^{\rm K} - N_{ij}^{\rm bkgnd}
                        - \sum_l
                                 \epsilon_l
                                 s_{i}^{l} S_{ij}^l
                                 G^l_{j}
                                 R
                            }
                          \over {\Delta N_{ij}^{\rm K}}} \right]^2
\nonumber \\
        & + & \sum_k  \left[
                       {{N_{k}^{\rm K} - N_{k}^{\rm bkgnd}
                        - \sum_l
                                 \epsilon_l
                                 T_{k}^l
                                  H^l_{k}
                                  R
                            }
                          \over {\Delta N_{k}^{\rm K}}} \right]^2,
\label{eqn:chi}
\end{eqnarray}
where $l$ is an index referring
to either \tk, \tkz, or \tkzz, and
the indices $i$, $j$, and $k$  run over the $p$,
$E/p$, and $N_{\rm clus}$ bins, respectively. The number
of background kaons within each bin is
$N^{\rm bkgnd}$,
while
$\Delta N^{\rm K}$ is the
statistical uncertainty on the number of
kaons within each bin, $N^{\rm K}$.
Also:
\begin{eqnarray}
 & R  &                 = {{N^{\rm K}_{\rm TOTAL}}
                     \over{f \ek + (1-f)\ekoz}}
\nonumber \\
 &   G_{j}^{\rm K^-} & =
                 {{\cal{S}}}_j^{\pi^-} f
\nonumber \\
 &  G_{j}^{\rm K^-h^0} & =
                 {{\cal{S}}}_j^{\pi^-\pi^0}
                 (1-f) {{\bkz} \over {\bkz+\bkzz}}
\nonumber \\
 &  G_{j}^{\rm K^-\ge2h^0} & =
                 {{\cal{S}}}_j^{\pi^-\pi^0\pi^0}
                 (1-f) {{\bkzz}\over {\bkz+\bkzz}}
\nonumber \\
 &   H_{k}^{\rm K^-} & =
                 {{\cal{T}}}_k^{\pi^-} f
\nonumber \\
 &  H_{k}^{\rm K^-h^0} & =
                 {{\cal{T}}}_k^{\pi^-\pi^0}
                 (1-f) {{\bkz} \over {\bkz+\bkzz}}
\nonumber \\
 &  H_{k}^{\rm K^-\ge2h^0} & =
                 {{\cal{T}}}_k^{\pi^-\pi^0\pi^0}
                 (1-f) {{\bkzz}\over {\bkz+\bkzz}},
\nonumber
\end{eqnarray}
and
\begin{description}
\item[$f$] is the ratio of the OPAL \tk\ branching ratio
to the OPAL \tknnz\ inclusive branching ratio, and is
allowed to float in the fit.
\item[$N^{\rm K}_{\rm TOTAL}$] is the 
total number of charged kaons in the one-prong tau
decay sample, as determined in Section~5, and shown in Table~\ref{tab:sum1b}.
\item[{\rm $\epsilon_l$}] is the efficiency for events from tau
decay $l$ in the tau-pair candidate sample to contribute
to the one-prong tau decay sample. The \ekz, \ekzz, and \ekoz\ efficiencies
are calculated from the average of the relevant efficiencies appearing in
Table~\ref{tab:br}, weighted by the associated world average branching
ratios.  All efficiencies have been corrected for biases introduced by
the tau-pair selection procedure.
\item[{\rm \bkz\ and \bkzz}] are the world average branching ratios
for \tkz\ and \tkzz\ decays, respectively, 
derived from results presented in
reference \cite{bib:pdg}.
\item[$S^l_{ij}$] is the probability, calculated
using Monte Carlo generated events, for tau decays of
type $l$ in the one-prong tau sample to contribute
to the $(i,j)^{\rm th}$
bin of the $(p,E/p)$ distribution.
\item[{\rm ${{\cal{S}}}_j^{\pi^-}$,
           ${{\cal{S}}}_j^{\pi^-\pi^0}$,
           and ${{\cal{S}}}_j^{\pi^-\pi^0\pi^0}$}]
are the corrections to the $j^{\rm th}$ bin of
the Monte Carlo $E/p$ distribution
of the various one-prong tau decays to pions,
as determined
from the $\chi^2$ fit using Equation \ref{eqn:chi1}.
\item[$s^l_{i}$] is a normalisation factor used to ensure that the
correction to $E/p$ does not affect the shape of the momentum distribution
of decays of type $l$.
\item[$T^l_{k}$] is the probability, calculated
using Monte Carlo generated events, for tau decays of
type $l$ in the one-prong tau sample to contribute
to the $k^{\rm th}$
bin of the $N_{\rm clus}$ distribution.
\item[{\rm ${{\cal{T}}}_j^{\pi^-}$,
           ${{\cal{T}}}_j^{\pi^-\pi^0}$,
           and ${{\cal{T}}}_j^{\pi^-\pi^0\pi^0}$}]
are the corrections to the $k^{\rm th}$ bin of
the Monte Carlo $N_{\rm clus}$ distribution
of the various one-prong tau decays to pions,
as determined in the previous section.
\end{description}

The value of $f$ derived from the result of the
fit using Equation~\ref{eqn:chi} is ${f=0.436\pm0.013}$,
where the uncertainty is statistical only.
The \tk\ branching ratio derived from this result is:
\begin{eqnarray}
\bk = {\rm Br}(\tk) & = & 0.658 \pm 0.024 \pm 0.029 \%,
\nonumber
\end{eqnarray}
where the first uncertainty is statistical and the second is
systematic, and the exclusive branching ratio is 60\% correlated
with the OPAL inclusive \tkzzz\ branching ratio.
A summary of the systematic uncertainties on \bk\ is
shown in Table~\ref{tab:sum2b}.

The statistical uncertainty in \bk\ due to the statistical
uncertainty in the number of kaons
within each bin of the $(p,E/p)$ and $N_{\rm clus}$
distributions
is assessed by repeating the fit
many times, each time varying the central
value of the number of kaons within a
particular bin, a bin at a time, by plus, then minus, one
statistical standard deviation.  The square root
of half of the quadrature sum of the net variations
produced in the branching ratio is taken as the statistical
uncertainty.

\begin{figure}[ht]
    \begin{center}
      \mbox{ \epsfxsize=15cm
            \epsffile[23 165 524 652]{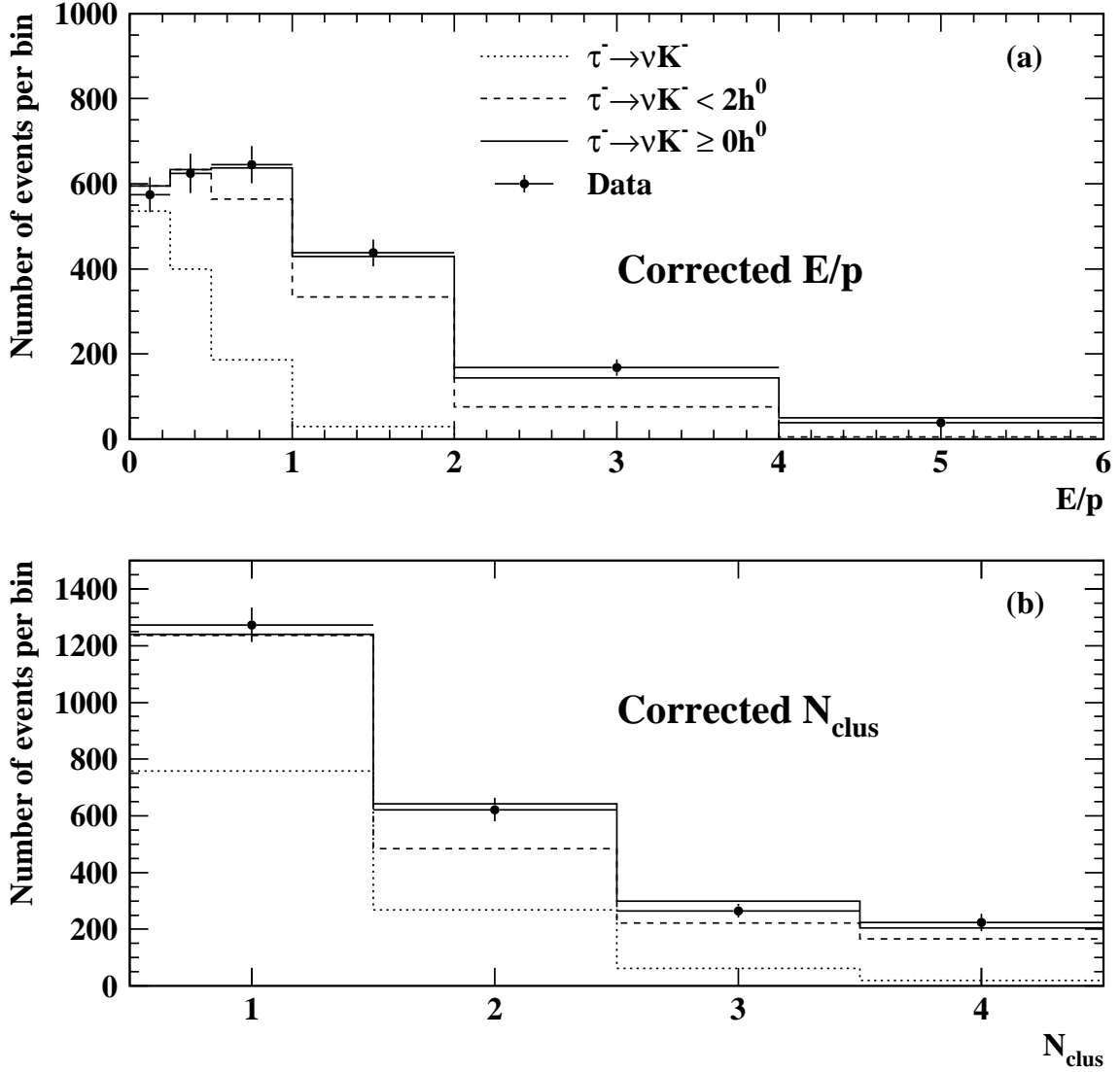} }
    \end{center}
  \caption[$E/p$ distribution of one-prong
tau decays with charged kaons]{
  \label{fig:probk}
(a) and (b) are the corrected
$E/p$ and $N_{\rm clus}$ distributions, respectively, of
Monte Carlo generated \tkzzz\ decays (histogram), along with the
$E/p$ and $N_{\rm clus}$ distributions of data one-prong tau decays with a charged
kaon in the final state (points).
The overall normalisation of Monte Carlo distributions is obtained from
the results of the \de\ likelihood
fit using Equation \ref{eqn:like1}, while
the normalisation of the \tk\ component is obtained from
the results of the $\chi^2$ fit using Equation \ref{eqn:chi}.
Backgrounds are on the order of $0.4\%$ or less and
are neglected in this plot.
The last bin in each of the distributions
contains overflow events.
  }
\end{figure}

The total $\chi^2$ per degree of freedom of the
fit using Equation \ref{eqn:chi}
is $63.0/50$, where $56.6/46$ results from the $\chi^2$ per
degree of freedom between the data and predicted $(p,E/p)$
distributions.
Figure \ref{fig:probk} shows the $E/p$ and $N_{\rm clus}$
distributions of one-prong
tau decays with a charged kaon, where the normalisation of
the contributions to the
predicted distribution comes from the OPAL \tk\ and \tknnz\
branching ratios.  If the $E/p$ and $N_{\rm clus}$ corrections
are not applied, the total $\chi^2$ per degree of freedom from the
fit is $79.9/50$.

To check for undue variation that may be produced by the choice
of binning used in the fit, the above procedure is repeated using various
binning schemes.
The RMS variation of the results returned by the
different fits is less than the statistical uncertainty from the
original fit.
In addition, a test of the entire fitting procedure
(including the calorimeter variable correction procedure
described in Sections~$6.1$ and $6.2$)
is performed with Monte Carlo generated
samples, and verifies that the resulting estimates of $f$  
are efficient and 
unbiased.

\subsection{\de\ Systematic Error}
\leavevmode\indent
The uncertainty in the $N^{\rm K}$ within each $N_{\rm clus}$ or $(p,E/p)$
bin
due to uncertainty in the
\de\ correction factors is determined as in Section~$5.2$.
The \de\ uncertainties are highly correlated between the bins.
Thus the uncertainty in \bk\ due to the \de\ systematic
uncertainty in the number of kaons
within each bin of the $(p,E/p)$ and $N_{\rm clus}$
distributions
is assessed by repeating the fit
many times,
each time varying the the central
value of the number of kaons within a
particular bin, a bin at a time, by plus, then minus, one
standard deviation of the \de\ systematic uncertainty.  Half of the linear
sum of the absolute variations
produced in the branching ratio is taken as the \de\ systematic
uncertainty in \bk,
and is shown in Table~\ref{tab:sum2b}.

\subsection{Systematic Uncertainty Associated with the Efficiency Estimation}
\leavevmode\indent
The uncertainty in the \tk\ branching ratio due
to the limited Monte Carlo statistics used to determine
the \ek, \ekz, and \ekzz\ efficiencies
is assessed
via the prior-belief method by modifying
Equation \ref{eqn:chi} such that:
\begin{eqnarray}
{\chi^2}^\prime = \chi^2
 + \left(
{{\epsilon_k^\prime - \epsilon_k}\over{\Delta\epsilon_k^\prime}}
\right)^2,
\nonumber
\end{eqnarray}
where
\begin{description}
\item[$\epsilon_k^\prime$ {\rm and}
$\Delta\epsilon_k^\prime$]
are the efficiency and associated uncertainty, due
to limited Monte Carlo statistics,
for kaons from signal channel $k$
in the tau-pair candidate sample to enter
the data sample.
\item[$\epsilon_k$] is the efficiency used in the
fit for kaons from
signal channel $k$ to contribute to the data sample.
\end{description}

In the first iteration of the fit the $\epsilon_k$ are allowed
to float.  In the second, they are
fixed to the values from the first iteration.  The
square root of the quadrature
difference in the errors in the branching ratio is taken
as the systematic uncertainty associated
with the limited Monte Carlo statistics used in
the assessment of the efficiencies, and is shown in
Table~\ref{tab:sum2b}.

Various world average exclusive branching ratios are used in the
calculation of \ekz\ and \ekzz\ (for instance, \ekz\ is the
average of \ekp\ and \ekk, weighted by the world average branching
ratios \bkp\ and \bkk).  In addition, these
same branching ratios are used to assess the
inclusive \tkz\ and \tkzz\
branching ratios
appearing in Equation \ref{eqn:chi}.
The uncertainty in
the \tk\ branching ratio due to the uncertainties in these
world average branching ratios is assessed 
by varying the branching ratios,
one at a time, by plus, then minus, one
standard deviation.  The square root
of half of the quadrature sum of the net variations
produced in the branching ratio is taken as the systematic
uncertainty, and is shown in Table~\ref{tab:sum2b}.

The systematic uncertainty in the efficiency and 
branching fraction due to the
uncertainty associated with the
Monte Carlo modelling of the probability for photons to convert to
an \epem\ pair
in the OPAL detector material is assessed as described in
Section~$5.3$.

\subsection{Systematic Uncertainty Associated with the Modelling of $E/p$ and
 $N_{\rm clus}$}
\leavevmode\indent
The uncertainty in the \tk\ branching ratio due to the
statistical uncertainties in the $E/p$ correction
factors is assessed using the prior-belief method by modifying
Equation \ref{eqn:chi} such that:
\begin{eqnarray}
{\chi^2}^\prime = \chi^2
+ {\bf (n^\prime-n)^T V^{-1} (n^\prime-n)}  ,
\nonumber
\end{eqnarray}
where
\begin{description}
\item[${\bf n^\prime}$] is a vector containing the central
values of the $E/p$ corrections, as
obtained from the fit to the \nptz\ sample.
\item[${\bf n}$] is a vector containing the
values of the $E/p$ corrections
used in the $\chi^2$ fit.
\item[${\bf V}$] is the covariance matrix
for the ${{\cal{S}}}$ correction factors, as obtained from the
fit to the \nptz\ $(p,E/p)$ and $N_{\rm clus}$ distributions.
\end{description}

In the
first iteration, all bins of the three sets of ${{\cal{S}}}$
corrections
are allowed to vary in the fit, and the returned values are
found to be consistent with the input values.
In the second iteration, the fit is repeated, this time keeping the
$E/p$ corrections
fixed to the values from the first iteration.  The
systematic uncertainty in \bk\ due to the
modelling of $E/p$
is then obtained from the
square root of the
quadrature difference of the fit uncertainties
from the two iterations.

The uncertainty in the \tk\ branching ratio due to the
statistical uncertainties in the $N_{\rm clus}$ correction
factors
is assessed in an analogous fashion.
The combined systematic uncertainty in \bk\ due to the corrections
to the modelling of $E/p$
and $N_{\rm clus}$
is shown in
Table~\ref{tab:sum2b}.

\section{Summary and Discussion}

From an analysis of the ionisation energy loss
of charged particles selected from
a sample of 220652 tau decay candidates recorded in the
barrel region of the OPAL
detector
at \epem\ centre-of-mass energies close to the Z$^0$ resonance,
we determine the branching ratios:
\begin{eqnarray*}
{\rm Br}(\tkzzz) & = & 1.528 \pm 0.039 \pm 0.040 \%
\nonumber \\[2mm]
{\rm Br}(\tk)    & = & 0.658 \pm 0.024 \pm 0.029 \%
\nonumber \\[2mm]
{\rm Br}(\tknz)  & = & 0.869 \pm 0.031 \pm 0.034 \%
\nonumber
\end{eqnarray*}
where the first uncertainties are statistical
and the second are systematic.  The \tk\ and \tknz\ 
branching ratios are
60\% and 80\% correlated to the \tkzzz\ 
branching ratio, respectively.
 
Both the inclusive and exclusive branching ratios are in
agreement with previous measurements,
shown in Table \ref{tab:tab1f}, and
both
contribute significantly to the reduction of the uncertainty
on the world averages of the respective quantities.

The theoretical prediction for the
\tk\ branching ratio is calculated via~\cite{bib:barish}:
\begin{eqnarray}
{\rm Br}(\tk)  & = &
{{G_{\rm F}^2 m_\tau^3 \tau_\tau} \over {16 \hbar \pi}}
f_{\rm K}^2 |{\rm V}_{us}|^2
\left( 1-m_{\rm K}^2/m_\tau^2 \right)^2 
\nonumber \\
               & = & 0.705 \pm 0.008 \%,
\label{eqn:predict1}
\end{eqnarray}
where the values of $f_{\rm K} = 0.1598\pm0.0015$~GeV and
$\vus=0.2196\pm0.0023$ are both taken 
from reference~\cite{bib:pdg}.
The OPAL measurement is in agreement with this prediction
to within $1.3$ standard deviations
of the combined uncertainties.
 
The OPAL measurement of the \tk\ branching fraction may be
used to determine either $f_{\rm K}$ or \vus\ through
Equation~\ref{eqn:predict1}.  
Assuming $f_{\rm K} = 0.1598\pm0.0015$~GeV,
a value of $\vus=0.2121\pm0.0063$ is obtained. 
Assuming
a value of $\vus=0.2196\pm0.0023$, a value of 
$f_{\rm K} = 0.1544\pm0.0046$~GeV is obtained.
In both cases, the uncertainty on the determined
quantity is dominated by the
uncertainty on the \tk\ branching fraction.
 
The theoretical prediction for the \tkzzz\ branching ratio is
obtained by summing the theoretical predictions for the
five states appearing in Table~\ref{tab:tab2b}:
\begin{eqnarray}
{\rm Br}(\tkzzz) & = & 1.571 \pm 0.043 \%.
\nonumber
\end{eqnarray}
The OPAL measurement is in agreement with this prediction to
within $0.7$ standard deviation of the combined uncertainties.

The OPAL collaboration has previously presented results
for the tau branching fractions to the
$\nu_\tau {\rm K}^- \pi^- \pi^+ \ge 0 \pi^0$,
$\nu_\tau \pi^- {\rm K}^0$, and
$\nu_\tau \pi^- {\rm K}^0 \ge 1 \pi^0$ final 
states \cite{bib:me,bib:sobie}.  These branching ratios
can be combined with the OPAL measurement of the
\tk\ branching ratio, using isospin relations appearing in \cite{bib:iso},
to yield an estimate of the branching fraction of the $\tau^-$ to
final states with strangeness $-1$:
\begin{eqnarray}
{\rm Br}(\tau^-\rightarrow [\mbox{\rm strangeness=-1}]) = 2.81\pm0.16\pm0.10 \%,
\nonumber
\end{eqnarray}
where the calculation of the uncertainties
takes into account all correlations between the various
OPAL results.
The calculation neglects contributions from \tketa\ to the final state,
and assumes that the
the \tkppzz\ final state is
dominated by \tkpp,
and that the 
${\tau^-\rightarrow \nu_\tau \pi^- {\rm K}^0 \ge 1 \pi^0}$
final state is dominated by 
${\tau^-\rightarrow \nu_\tau \pi^- {\rm K}^0 \pi^0}$.
The OPAL result is in agreement with the previously 
published result~\cite{bib:naleph2},
${\rm Br}(\tau^-\rightarrow [\mbox{\rm strangeness=-1}]) = 2.87\pm0.12 \%$,
to within $0.3\sigma$ of the combined uncertainties.
 
\appendix
\par
{\Large {\bf Acknowledgements}}
\par
We particularly wish to thank the SL Division for the efficient operation
of the LEP accelerator at all energies
 and for their continuing close cooperation with
our experimental group.  We thank our colleagues from CEA, DAPNIA/SPP,
CE-Saclay for their efforts over the years on the time-of-flight and trigger
systems which we continue to use.  In addition to the support staff at our own
institutions we are pleased to acknowledge the  \\
Department of Energy, USA, \\
National Science Foundation, USA, \\
Particle Physics and Astronomy Research Council, UK, \\
Natural Sciences and Engineering Research Council, Canada, \\
Israel Science Foundation, administered by the Israel
Academy of Science and Humanities, \\
Minerva Gesellschaft, \\
Benoziyo Center for High Energy Physics,\\
Japanese Ministry of Education, Science and Culture (the
Monbusho) and a grant under the Monbusho International
Science Research Program,\\
Japanese Society for the Promotion of Science (JSPS),\\
German Israeli Bi-national Science Foundation (GIF), \\
Bundesministerium f\"ur Bildung und Forschung, Germany, \\
National Research Council of Canada, \\
Research Corporation, USA,\\
Hungarian Foundation for Scientific Research, OTKA T-029328,
T023793 and OTKA F-023259.\\


\clearpage


\begin{table}
\begin{center}
\begin{tabular}{|l|l|} \hline
 $N_\tau$                   & 220652            \\
 $f^{{\rm non-}\tau}$       & $1.56\pm0.10\%$   \\
\# in one-prong sample      & $143528$           \\ \hline
 $N_{\rm TOTAL}^{\pi}$      & $75566\pm389\pm246$  \\
 $N^{\pi}_{\rm bkgnd}$      & $89\pm16$    \\
\epnz\                      & $0.720\pm0.001\pm0.001$  \\ \hline
 $N_{\rm TOTAL}^{\rm K}$    & $2526\pm64\pm52$  \\
 $N^{\rm K}_{\rm bkgnd}$    & $9\pm4$    \\
\eknz\                      & $0.759\pm0.006\pm0.007$  \\
\hline
\end{tabular}
\end{center}
\caption[Hadron composition of the one-prong tau decay sample]{
Pion and kaon composition of the one-prong
tau decay candidate
sample, as estimated by the likelihood fit to the
measured \de\ of tracks in the sample.  The first
uncertainty on
$N^\pi_{\rm TOTAL}$
and
$N^{\rm K}_{\rm TOTAL}$
is the statistical uncertainty from the
fit, and the second is due to the systematic uncertainty
in the \de\ correction factors.
Also shown are the estimated backgrounds in $N^{\pi}_{\rm TOTAL}$ and
$N^{\rm K}_{\rm TOTAL}$,
along with the
efficiencies for signal events to contribute
to the one-prong tau decay sample.
The first uncertainty on each efficiency estimate
is due to the limited statistics of the Monte Carlo generated
samples, while the second is due to the uncertainty in the
various branching ratios used to calculate the efficiency, as listed
in Table~\ref{tab:br}.
}
\label{tab:sum1b}
\end{table}

\begin{table}
\begin{center}
\begin{tabular}{|l|c|c|} \hline
 $\tau^-$ Decay & BR (\%)         & Efficiency  \\
 Mode           &                 &             \\
\hline
\np       & $11.08\pm0.13$         & $0.810\pm0.003$  \\
\nppz     & $25.32\pm0.15$         & $0.725\pm0.002$  \\
\npkz     & $0.83\pm0.08$          & $0.604\pm0.007$  \\
\npppz    & $9.15\pm0.15$          & $0.642\pm0.003$  \\
\tpkkz    & $0.121\pm0.021$        & $0.454\pm0.016$  \\
\tpkpz    & $0.39\pm0.05$          & $0.550\pm0.015$  \\
\nppppz   & $1.11\pm0.14$          & $0.569\pm0.005$  \\         
\npeta    & $0.188\pm0.025$        & $0.655\pm0.024$  \\ \hline
\tk       & -                      & $0.843\pm0.010$  \\
\tkpz     & $0.52\pm0.05$          & $0.757\pm0.011$  \\
\tkkz     & $0.159\pm0.024$        & $0.661\pm0.013$  \\
\tketa    & $0.027\pm0.006$        & $0.757\pm0.011$  \\
\nkppz    & $0.080\pm0.027$        & $0.664\pm0.025$  \\
\nkkpz    & $0.151\pm0.029$        & $0.544\pm0.012$  \\
\tkpppz   & $0.035\pm0.024$        & $0.643\pm0.025$  \\
\tkkppz   & $0.000\pm0.018$        & $0.560\pm0.012$  \\
\hline
\end{tabular}
\end{center}
\caption[Branching ratios and efficiencies used to extract the \tknnz\
and \tpnnz\
branching ratios]{
\label{tab:br}
Branching ratios and efficiencies used to estimate the
efficiencies for the \tknnz\ and \tpnnz\ decays in the
tau-pair sample to contribute to the one-prong
sample.
The \tkkppz\ branching fraction is estimated from the result
appearing in \cite{bib:naleph}.
}
\end{table}


\begin{table}
\begin{center}
\begin{tabular}{|l|r|r|r|} \hline
 & \multicolumn{3}{c|}{Branching Ratios ($\%$)} \\ \cline{2-4}
\multicolumn{1}{|c|}{}
                     & \bpnz & \bknz      & \bk
\\ \hline
\rule[-1mm]{0cm}{6mm}
Central value        &
\multicolumn{1}{r}   { $48.17$}       &
\multicolumn{1}{|r}  { $1.528$}        &
\multicolumn{1}{|r|} { $0.658$} \\ \hline
$\sigma$ (stat)
                     & { ${\pm 0.25}$}
                     & { ${\pm 0.039}$}
                     & { ${\pm 0.024}$} \\
$\sigma$ (\de\ sys)
                     & { ${\pm 0.16}$}
                     & { ${\pm 0.032}$}
                     & { ${\pm 0.017}$} \\
$\sigma$ ($E/p$ and $N_{\rm clus}$ modelling)
                     &  -
                     &  -
                     & { ${\pm 0.014}$} \\
$\sigma$ (photon conversion modelling)
                     & { ${\pm 0.68}$}
                     & { ${\pm 0.015}$}
                     & { ${\pm 0.003}$} \\
$\sigma$ (MC stat)
                     & { ${\pm 0.09}$}
                     & { ${\pm 0.012}$}
                     & { ${\pm 0.006}$} \\
$\sigma$ (efficiency sys)
                     & { ${\pm 0.05}$}
                     & { ${\pm 0.014}$}
                     & { ${\pm 0.017}$} \\
\hline
\end{tabular}
\end{center}
\caption[One-prong inclusive branching ratio and sources of systematic
uncertainty]{
Summary of the branching ratio
central values and sources of uncertainties.
The second-to-last 
uncertainty is the systematic uncertainty
arising from the limited statistics of the Monte Carlo generated samples
used to estimate the selection efficiencies.
The last uncertainty is due to the systematic uncertainty in the efficiency
correction arising from the uncertainties in the various world average
branching
ratios used in the weighted average calculation of the
\tknnz, \tpnnz, and \tknz\
efficiencies.
}
\label{tab:sum2b}
\end{table}


\begin{table}
\begin{center}
\begin{tabular}{|l|cl|} \hline
 \multicolumn{1}{|c|}{$\tau^-$ DECAY} &
 \multicolumn{2}{c|}{THEORY}          
  \\
 \multicolumn{1}{|c|}{MODE}          &
 \multicolumn{2}{c|}{BR $(\%)$}      
  \\ \hline
\tk     &  $0.705\pm0.008$ & \cite{bib:barish}
         \\ \hline
\tkpz   &  $0.48\pm0.02$   & \cite{bib:barish}
         \\ \hline
\tkkz   &  $0.111\pm0.031$ & \cite{bib:CVC2}
         \\ \hline
\nkppz  &  $0.12\pm0.02$   & \cite{bib:theory2}
         \\ \hline
\nkkpz  & $0.155\pm0.005$  & \cite{bib:theory2}
         \\ \hline
\end{tabular}
\end{center}
\caption[foo]{
\label{tab:tab2b}
Theoretical predictions for tau branching ratios
to the dominant one-prong final states that include a
charged kaon.}
\end{table}

\begin{table}
\begin{center}
\begin{tabular}{|l|ll|} \hline
\multicolumn{1}{|c|}{$\tau^-$ DECAY} & \multicolumn{2}{c|}{EXPERIMENT}  \\ 
\multicolumn{1}{|c|}{MODE}           & \multicolumn{2}{c|}{BR $(\%)$}   \\ \hline
\rule[-1mm]{0cm}{6mm}
\tk  & $0.696\pm0.025\pm0.014$ & ALEPH99\cite{bib:naleph} \\
     & $0.85\pm0.18$ & DELPHI94\cite{bib:ndelphi} \\
     & $0.66\pm0.07\pm0.09$ & CLEO94\cite{bib:ncleo}     \\
     & $0.59\pm0.18$ & DELCO84\cite{bib:ndelco}   \\ 
     & $0.658\pm0.024\pm0.029$ & (this analysis)          \\ \hline
\tknnz  & $1.52\pm0.04\pm0.04$ & ALEPH99\cite{bib:naleph}  \\
        & $1.54\pm0.24$ & DELPHI94\cite{bib:ndelphi}\\
        & $1.60\pm0.12\pm0.19$ & CLEO94\cite{bib:ncleo}    \\
        & $1.71\pm0.29$ & DELCO84\cite{bib:ndelco}  \\ 
        & $1.528\pm0.039\pm0.040$ & (this analysis)         \\ \hline
\end{tabular}
\end{center}
\caption[foo]{
\label{tab:tab1f}
Summary of the \tk\ and \tknnz\ branching ratios.
}
\end{table}


\end{document}